%% file: ms.tex
\documentclass[journal]{IEEEtran}
\usepackage{amsmath,amsfonts,amssymb}
\usepackage{algorithm}
\usepackage{algpseudocode}  
\usepackage{array}
\usepackage{subcaption}
\usepackage{pifont}
\usepackage{textcomp}
\usepackage{stfloats}
\usepackage{url}
\usepackage{verbatim}
\usepackage{graphicx}
\usepackage{listings}
\usepackage{cite}
\usepackage{siunitx}
\usepackage{tikz}
\usepackage[normalem]{ulem}
\usepackage{cancel}
\usepackage{balance}
\input{macros.tex}

\usepackage{comment}
\includecomment{supplement}

\usepackage{cuted}
\usepackage{capt-of}

\usepackage{hyperref}

\begin{document}

\title{Approximating Analytically-Intractable Likelihood Densities with
Deterministic Arithmetic for Optimal Particle Filtering}

\author{Orestis Kaparounakis, Yunqi Zhang, and Phillip Stanley-Marbell%
\thanks{The authors are with the Department of Engineering, University of
Cambridge, CB3 0FA UK (e-mail: ok302@cam.ac.uk; yz795@cam.ac.uk;
psm751@cam.ac.uk). Phillip Stanley-Marbell is also with
Signaloid Ltd.}
\thanks{The research results presented in the article are part of commercial
activity at Signaloid in which Orestis Kaparounakis and Phillip Stanley-Marbell
have a commercial interest.}
\thanks{This work was supported by the EPSRC.}
\thanks{This version was accepted for publication at IEEE Signal Processing
Letters. DOI: \url{https://doi.org/10.1109/LSP.2026.3664784}.}}

\maketitle

\input{text/abstract.tex}

\begin{IEEEkeywords}
Particle filtering, non-Gaussian non-linear observation models, optimal
likelihood, native uncertainty-tracking, state estimation.
\end{IEEEkeywords}

\input{text/introduction.tex}

\input{text/core.tex}

\input{text/results.tex}

\input{text/conclusion.tex}

\bibliographystyle{IEEEtran}
\bibliography{references}

\begin{supplement}
\clearpage
\pagebreak
\input{text/supplementary.tex}
\end{supplement}

\end{document}

%% file: macros.tex
\usepackage{color}
\usepackage[dvipsnames, table]{xcolor}

\usepackage[sfdefault]{libertine}      
\usepackage{libertinust1math}          
\usepackage[scaled=0.8]{beramono}      
\usepackage[T1]{fontenc}

\newcommand\pdfs     {\textsc{\large pdf}s}

\newcommand\codelight[1]        {{\ttfamily #1}}

\def\cameraready{}

\ifdefined\cameraready
	\newcommand\remove[1]	{}
	
\else
	\newcommand\remove[1]	{\textcolor{red}{#1}}
	
\fi

\newcommand\rmtext[1]		{}

\newcommand{\stkout}[1]{\ifmmode\text{\sout{\ensuremath{#1}}}\else\sout{#1}\fi}

\makeatletter
\newcommand{\suppref}[2]{%
  \ifcsname r@#1\endcsname
    \ref{#1}%
  \else
    #2%
  \fi
}
\makeatother

\definecolor{a}{rgb}{0.9,0.95,0.95}
\definecolor{b}{rgb}{0.99,0.99,0.99}

\definecolor{C0}{HTML}{1F77B4} 
\definecolor{C1}{HTML}{FF7F0E} 
\definecolor{C2}{HTML}{2CA02C} 

\definecolor{comment}{RGB}{0,128,0} 
\definecolor{string}{RGB}{255,0,0}  
\definecolor{keyword}{RGB}{0,0,255} 

\definecolor{OmnigraffleLightMocha}{rgb}{0.702, 0.267, 0.0}

\definecolor{babyblue}{rgb}{0.54, 0.81, 0.94}
\definecolor{babyblueeyes}{rgb}{0.63, 0.79, 0.95}
\definecolor{ballblue}{rgb}{0.13, 0.67, 0.8}
\definecolor{asparagus}{rgb}{0.53, 0.66, 0.42}
\definecolor{amethyst}{rgb}{0.6, 0.4, 0.8}

\definecolor{listinggreen}{rgb}{0,0.6,0}
\definecolor{listinggray}{rgb}{0.5,0.5,0.5}
\definecolor{listingmauve}{rgb}{0.58,0,0.82}
\definecolor{listingkeywordcolor}{rgb}{1.0,0.4,0.0}
\definecolor{listinglightgray}{rgb}{0.9863,0.9863,0.9863}

\lstdefinestyle{c}{
    backgroundcolor=\color{listinglightgray},
	commentstyle=\color{comment},
	stringstyle=\color{string},
	keywordstyle=\color{amethyst},
	basicstyle=\footnotesize\ttfamily,
	numbers=left,
	numberstyle=\tiny,
	numbersep=5pt,
	frame=lines,
	breaklines=true,
	prebreak=\raisebox{0ex}[0ex][0ex]{\ensuremath{\hookleftarrow}},
	showstringspaces=false,
	upquote=true,
	tabsize=2,
	deletekeywords = {double},
	keywordstyle = [2]\color{asparagus},
	morekeywords = [2]{double, int, size_t},
    emphstyle = [1]\color{ballblue},
    emph      = [1]{
		UxHwDoubleSample,
		sample,
		observation_model_ideal,
		observation_noise_model,
		evaluate_likelihood_analytical,
		evaluate_density_analytical,
		evaluate_likelihood_uxhw,
		evaluate_density_uxhw,
		transition_model_ideal,
		transition_noise_model},
}

%% file: text/abstract.tex
\begin{abstract}
Particle filtering algorithms have enabled practical solutions to problems in
autonomous robotics (self-driving cars, UAVs, warehouse robots), target
tracking, and econometrics, with further applications in speech processing and
medicine (patient monitoring). Yet, their inherent weakness at representing the
likelihood of the observation (which often leads to particle degeneracy) remains
unaddressed for real-time resource-constrained systems. Improvements
such as the optimal proposal and auxiliary particle filter mitigate this issue
under specific circumstances and with increased computational cost. This work
presents a new particle filtering method and its implementation, which enables
tunably-approximative representation of arbitrary likelihood densities as
program transformations of parametric distributions.
Our method leverages a recent computing platform that
can perform deterministic computation on probability distribution
representations (UxHw) without relying on stochastic methods.
For non-Gaussian non-linear systems and with an optimal-auxiliary particle
filter, we benchmark the likelihood evaluation error and speed for a total of
294\,840 evaluation points. For such models, the results show that the UxHw
method leads to as much as 37.7x speedup compared to the Monte Carlo
alternative. For narrow uniform measurement uncertainty, the particle filter falsely
assigns zero likelihood as much as 81.89\% of the time whereas UxHw achieves
1.52\% false-zero rate. The UxHw approach achieves filter RMSE improvement of as
much as 18.9\% (average 3.3\%) over the Monte Carlo alternative.
\end{abstract}

%% file: text/introduction.tex
\section{Introduction}
\IEEEPARstart{P}{article} filters~\cite{gordon1993novel} have become a mainstay
for real-time state estimation in embedded systems because they handle nonlinear
dynamics and non-Gaussian noise better than classical Kalman filters and
variants~\cite{kalman1961new, lefebvre2004kalman}. Yet, the accuracy and
performance of particle filters hinge on an explicit likelihood
model~\cite{liu1998sequential},
which in sensor-rich, data-driven environments is often unavailable, intractable
to formulate~\cite{toni2009approximate, sisson2007sequential,
sigges2017likelihood}, or dynamically changing~\cite{ren2024line}.
Computing the likelihood for analytically-intractable models can be
computationally expensive~\cite{alali2024kernel,
fengler2021likelihood}. These computations are ill-suited for
resource-constrained embedded platforms under tight power and latency budgets.

This letter introduces a new method for particle filter implementation that
enables easy marginalization over noise distributions, facilitating the
computation of the likelihood density. The approach capitalizes on recent
advances in hardware architectures for performing arithmetic on digital
representations of probability density functions
(\pdfs{})~\cite{tsoutsouras2022laplace,bilgin2025quantization,petangoda2025monte}.

Particle filters compute the likelihood density at the measurement to adjust the
particle weights and thus update the state estimate. When the likelihood is hard
to compute, this step becomes a bottleneck for accurate and performant state
estimation~\cite{li2016numerical}.
This article introduces a technique for approximating the measurement likelihood
for analytically-intractable probability distributions. The technique leverages
direct arithmetic on probability density functions and density evaluation via
uncertainty-extended hardware
(UxHw)~\cite{tsoutsouras2022laplace}. Here, we apply
the technique for computing the optimal likelihood for the auxiliary particle
filter~\cite{pitt1999filtering}.
Supplementary Section~\suppref{SuppSectionParticleFilteringRecap}{V-A} provides a 
brief refresher on particle filters.

This letter makes the following contributions to the state of the art:
\begin{enumerate}
    \item[\ding{192}] A tunably-approximative approach to computing likelihoods
    from arbitrary densities in particle filters. This method uses
    Turing-complete transformations on random variables in conventional
    programming language syntax. Building on functionality enabled by UxHw, it
    keeps likelihood models simple, digestible, and with clear semantic mapping
    to the sensor physics. The approach offers deterministic time guarantees,
    paving the way for use in real-time systems, without having to rely on
    complex Monte Carlo methods. (Section~\ref{Section_PLAPF})
    \item[\ding{193}] Experimental validation of the accuracy,
    bare-metal speed, robustness, and convergence, of the
    proposed method for the optimal likelihood on the Gordon--Salmond--Smith
    system~\cite{gordon1993novel}, across 540 system configurations including
    Gaussian and non-Gaussian noise, for a total of \qty{294840} evaluation
    points. (Section~\ref{Section_Results})
\end{enumerate}

The results show that the UxHw method enables Bayesian filtering for
non-Gaussian non-linear analytically-intractable models on
resource-constrained systems, such as drones and autonomous robots, as much as
37.7x faster than the status quo. This paves the way for real-time use of powerful likelihood models.

\lstset{escapeinside={(*@}{@*)}}

\begin{figure}[htb]
\centering
\begin{lstlisting}[language=C,style=c,caption=Baseline pointwise approach.,label=ListingProxyLyksPointwiseApproach]
double lyk_proxy_pointwise(double z, double x, double t) {
	double expected_x = transition_model_ideal(x, t);
	double expected_z = observation_model_ideal(expected_x);
	return evaluate_density_analytical(z, expected_z);
}
\end{lstlisting}
\hfill

\begin{lstlisting}[language=C,style=c,caption=Monte Carlo simulation.,label=ListingProxyLyksMonteCarloApproach]
double lyk_proxy_mc(double z, double x, double t) {
	double	accum = 0.0;
	double	expected_x = transition_model_ideal(x, t);
	for (size_t i = 0; i < M; i++) {
		double sim_noise = sample(transition_noise_model());
		double sim_x = expected_x + sim_noise;
		double sim_expected_z = observation_model_ideal(sim_x);
		double lk = evaluate_density_analytical(z, sim_expected_z);
		accum += lk;
	}
	return accum / (double)M;  // Return average
}
\end{lstlisting}
\hfill
\begin{lstlisting}[language=C,style=c,caption=UxHw approach.,label=ListingProxyLyksUxHwMarginalApproach]
double lyk_proxy_uxhw(double z, double x, double t) {  
	double x_density = transition_model_ideal(x, t)  (*@\tikz[overlay, baseline=-0.55ex]{\node[fill=ForestGreen!30,inner sep=2.5pt](A){+};\draw[densely dashed, line width=0.6pt](A.east)--++(0.45,0.25)node[anchor=west,font=\small]{\ding{202}};}@*)
	                   transition_noise_model();
	double z_density = observation_model_ideal(x_density)  (*@\tikz[overlay, baseline=-0.55ex]{\node[fill=ForestGreen!30,inner sep=2.5pt](A){+};\draw[densely dashed, line width=0.6pt](A.north east)--++(0.4,0.3)node[anchor=south west,font=\small]{\ding{202}};}@*)
	                   observation_noise_model();
	return            (*@\tikz[overlay, baseline=-0.65ex]{\node[fill=ForestGreen!22,inner sep=2pt](A){UxHwDoubleEvaluatePDF};\draw[densely dashed, line width=0.6pt](A.north west)--++(-0.75,0.1)node[anchor=east,font=\small]{\ding{203}};}@*)           (z, z_density);
}
\end{lstlisting}
\caption{C code for three approaches of the proxy likelihood evaluation:
pointwise (Listing~\ref{ListingProxyLyksPointwiseApproach}), Monte Carlo
simulation (Listing~\ref{ListingProxyLyksMonteCarloApproach}) , and UxHw-based
computation (Listing~\ref{ListingProxyLyksUxHwMarginalApproach}). Supplementary
Listing~\suppref{ListingComplementaryCode}{4} contains supplementary code definitions.
\ding{202} On an uncertainty-tracking processor~\cite{tsoutsouras2022laplace}
such as the one in Figure~\ref{fig:NativeUncertaintyTrackingHardware}, the
\colorbox{ForestGreen!30}{\codelight{+}} operator adds the conventional values
but also performs addition between the associated distributions for these
values. \ding{203}
\colorbox{ForestGreen!18}{\codelight{UxHwDoubleEvaluatePDF}} is not a C
function but rather a function-wrapped UxHw-microarchitecture
instruction for evaluating the density~\cite{bilgin2025evaluate}.}
\label{FigCodeComparePointwiseMonteCarloUxHw}
\end{figure}

\newcommand\textredsquare[0] {\textcolor{OrangeRed}{$\blacksquare$}}
\newcommand\textgreensquare[0] {\textcolor{ForestGreen}{$\blacksquare$}}

%% file: text/core.tex
\section{Predictive-Lookahead Auxiliary Particle Filter}
\label{Section_PLAPF}

Let $x_k$ be the state of a one-dimensional Markov system at time $k$ and
let $z_k$ denote the observation. Let $\upsilon_k\sim\mu_\upsilon$ and
$\nu_k\sim\mu_\nu$ denote transition and observation noise respectively, with
arbitrary distributions. Equations~\ref{EqTransitionGeneric}
and~\ref{EqObservationGeneric} are the state-space transition and observation
models for the system:
\begin{align}
    x_k&=f(x_{k-1},\upsilon_k), \label{EqTransitionGeneric}\\
    z_k&=h(x_k,\nu_k).   \label{EqObservationGeneric}
\end{align}

Let superscript $(i)$ denote the particle index. The \emph{auxiliary}
particle filter selects parent particles using \emph{lookahead weights}
$\tilde{w}_k^{(i)} \propto w_{k-1}^{(i)}\, m_k^{(i)}$, where $m_k^{(i)}$ is a
\emph{proxy likelihood}~\cite{pitt1999filtering, arulampalam2002tutorial}. Let
$X_k^{(i)}$ denote the next-state value of the $i$th-particle as a random
variable. A common choice evaluates the likelihood at a point prediction:
$m_k^{(i)} = p(z_k \mid \mathbb{E}_{\mu_\upsilon}[X_k^{(i)}])$.

A better choice is the \emph{predictive likelihood} in
Equation~\ref{EqParticleFilterUpdateOptimalGenericDensityShort} which accounts
for transition uncertainty~\cite{doucet2000sequential}:
\begin{align}
m_k^{(i)} &= p(z_k \mid x_{k-1}^{(i)}), \;\text{where}\nonumber\\
p(z_k \mid x_{k-1}^{(i)}) &= \int p(z_k \mid x_k)\, p(x_k \mid x_{k-1}^{(i)})\, dx_k.
\label{EqParticleFilterUpdateOptimalGenericDensityShort}
\end{align}
This choice avoids conditioning on a single point prediction of the transition.

In general, Equation~\ref{EqParticleFilterUpdateOptimalGenericDensityShort}
does not have an analytic solution, so implementations approximate it. A common
baseline estimates $m_k^{(i)}$ by Monte Carlo simulation of the transition and
observation models (Listing~\ref{ListingProxyLyksMonteCarloApproach}), which is
compute-intensive and adds an extra stochastic component to the filter.

Section~\ref{SectionApproximatingHardLikelihoods} leverages UxHw to compute
$m_k^{(i)}$ deterministically by propagating $\mu_\upsilon$ and $\mu_\nu$
through $f$ and $h$ and evaluating the resulting density at the observation $z_k$.

\subsection{Predictive
likelihood evaluation with deterministic distribution arithmetic}
\label{SectionApproximatingHardLikelihoods}

\newcommand{\tightcolorbox}[2]{%
  \begingroup
  \setlength{\fboxsep}{1.6pt}
  \colorbox{#1}{%
    \rule[-0.2ex]{-0.8pt}{1ex}
    #2%
  }%
  \endgroup
}

\begin{figure*}[tb]
    \centering
    \includegraphics[width=0.97\textwidth]{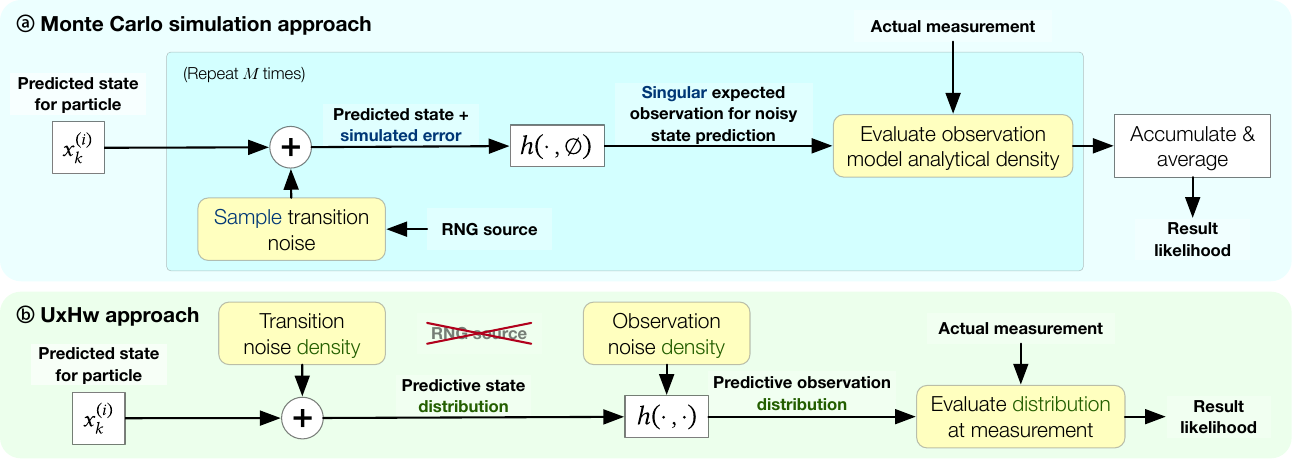}
    \caption{\textbf{\textcircled{\small a}} Diagram for the Monte Carlo simulation approach
    (Listing~\ref{ListingProxyLyksMonteCarloApproach}). $M$ independent re-executions \tightcolorbox{cyan!20}{sample}
    the transition noise and evaluate the analytic observation noise density at $h(f(x, \upsilon), 0)$, where
    $\upsilon\sim\mu_\upsilon$. The estimate for the proxy likelihood $m^{(i)}_k$ is the average of these evaluations.
    \textbf{\textcircled{\small b}} Diagram for the UxHw approach (Listing~\ref{ListingProxyLyksUxHwMarginalApproach}).
    In a single execution of the code, the UxHw arithmetic logic unit adds the entire transition noise density to the
    predicted state (Listing~\ref{ListingProxyLyksUxHwMarginalApproach} lines 2--3). Then, it pushes this predictive state \tightcolorbox{C2!20}{distribution}, together with the entire
    observation noise density, through the observation model $h$, computing the predictive observation
    \tightcolorbox{C2!20}{distribution} (Listing~\ref{ListingProxyLyksUxHwMarginalApproach} lines 4--5). The $m^{(i)}_k$
    estimate is the evaluation of the predictive observation \tightcolorbox{C2!20}{distribution} at the
    measurement (Listing~\ref{ListingProxyLyksUxHwMarginalApproach} line 6).}
    \label{LykProxyComputationGraph}
\end{figure*}

We approximate the predictive likelihoods $m_k^{(i)} = p(z_k \mid
x_{k-1}^{(i)})$
(Equation~\ref{EqParticleFilterUpdateOptimalGenericDensityShort}) using UxHw
which implements computation with deterministic arithmetic on digital
representations of probability density functions~\cite{bilgin2025quantization}.
This approach avoids sampling and instead composes transition and observation
uncertainty: the transition model outputs a distribution of predicted future
states, and supplying this distribution into the observation model yields the
\emph{predictive observation distribution}. Evaluating that distribution at the
measurement yields the predictive likelihood~\cite{bilgin2025evaluate}. The
Listings of Figure~\ref{FigCodeComparePointwiseMonteCarloUxHw} contains the
source code for the baseline pointwise approach
(Listing~\ref{ListingProxyLyksPointwiseApproach}), the Monte Carlo simulation
(Listing~\ref{ListingProxyLyksMonteCarloApproach}), and UxHw-based computation
(Listing~\ref{ListingProxyLyksUxHwMarginalApproach}). Supplementary
Listing~\suppref{ListingComplementaryCode}{4} contains example supplemental function
definitions for one system configuration.

Figure~\ref{LykProxyComputationGraph} shows the data flow for the Monte
Carlo and UxHw approaches. In the UxHw approach, for each particle
$x_{k-1}^{(i)}$, the code in Listing~\ref{ListingProxyLyksUxHwMarginalApproach} natively:
\begin{enumerate}
    \item computes the distribution of the random variable $X_k^{(i)} = f(x_{k-1}^{(i)}, \upsilon_k)$ in lines 2--3, then
    \item computes the induced distribution of $Z_k^{(i)} = h(X_k^{(i)}, \nu_k)$ in lines 4--5, and
    \item evaluates its density at the measurement $z_k$ to compute $\tilde m_{k,\eta}^{(i)} \approx p(z_k \mid x_{k-1}^{(i)})$ in line 6,
\end{enumerate} 
  where $\eta$ is a
configuration parameter of the UxHw processor, controlling the processor speed
versus distribution arithmetic fidelity tradeoff. This approximation has
predictable runtime and memory requirements suitable for real-time systems.

%% file: text/results.tex
\section{Results}
\label{Section_Results}

In practical state estimation scenarios, measurement models frequently yield
analytically-intractable likelihoods, due to their nonlinear, non-Gaussian, or
discontinuous structure~\cite{blanco2010optimal, cappe2007overview,
doucet2000sequential, li2021state}. We benchmark the accuracy and performance of
the UxHw method in the predictive-lookahead auxiliary particle filter on the
non-linear Gordon--Salmond--Smith state-transition and observation
models~\cite{gordon1993novel}, which Equations~\ref{EqGordonWalkTransition}
and~\ref{EqGordonWalkObservation} repeat:
\begin{align}
    x_k &= \frac{x_{k-1}}{2} + \frac{25 x_{k-1}}{1 + x_{k-1}^2} + 8 \cos(1.2 k) + \upsilon_k, \label{EqGordonWalkTransition}\\
    z_k &= \frac{x_k^2}{20} + \nu_k. \label{EqGordonWalkObservation}
\end{align}
The composition in Equation~\ref{EqParticleFilterUpdateOptimalGenericDensityShort} for the
transition and observation models of Equations~\ref{EqGordonWalkTransition}
and~\ref{EqGordonWalkObservation} does not have a closed-form solution, even
when the involved random variables follow Gaussian distributions.

For this system, we benchmark the computation of the likelihood density with the UxHw method
against: \ding{202} Monte Carlo estimation via the
predicitve likelihood, \ding{203} parametric analytic density evaluation via
variance approximation (i.e., EKF-style
linearization~\cite{johansen2009tutorial}), and \ding{204} the baseline
pointwise evaluation. 
We run the stochastic system under additive uncertainty for combinations of
$\mu_\upsilon$ and $\mu_\nu$ being Gaussian ($\mathcal{N}$), Laplacian
($\mathcal{L}$), and Uniform ($\mathcal{U}$), and for different scale
parameters. Supplementary Section~\suppref{SecLikelihoodBenchmarkingDetails}{V-B} fully
details the evaluation configuration.

\subsubsection*{Likelihood accuracy}
The UxHw approach enables systems to have better accuracy than the baseline
method, comparable with the Monte Carlo approach.
Figure~\ref{FigLikelihoodsErrorsECDF}\textsc{\large a} shows the empirical cumulative density
function (eCDF) for absolute errors of the different methods. UxHw consistently
achieves lower errors than the respective equal-speed Monte Carlo.

\subsubsection*{Filtering accuracy and sample effectiveness}
\begin{figure}[tb]
    \centering
    \includegraphics[trim={0.1cm 0.1cm 0.1cm 0.1cm},clip,width=\columnwidth]{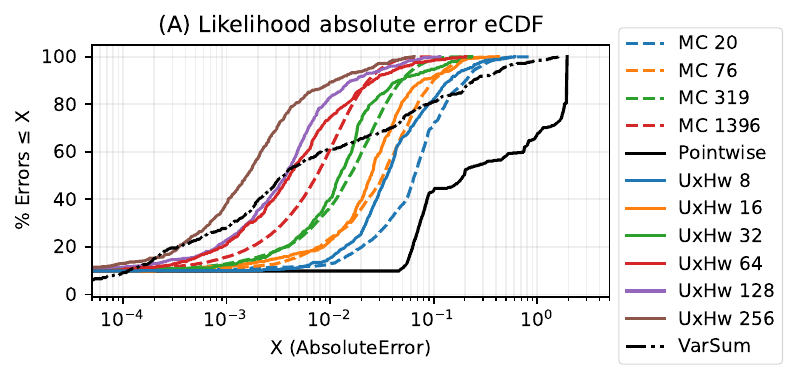}
    \hfill
    \includegraphics[trim={0.1cm 0.1cm 0.1cm 0.1cm},clip,width=\columnwidth]{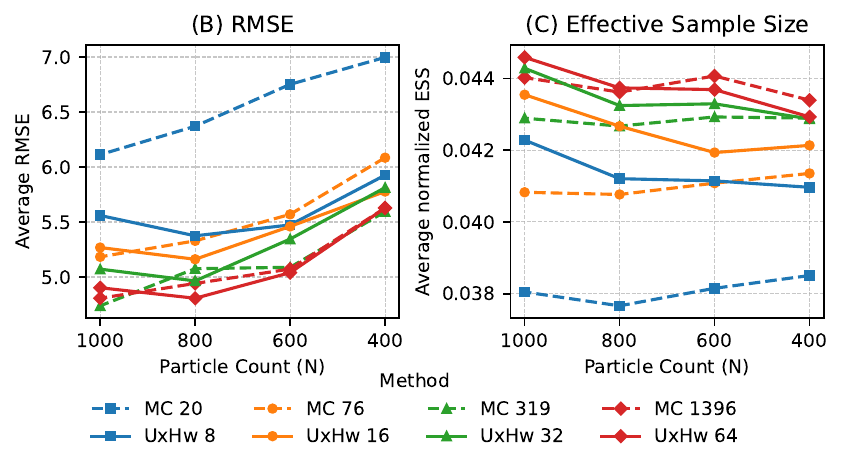}
    \caption{\textbf{(A)} \label{FigLikelihoodsErrorsECDF} Empirical
    distribution of absolute errors for each method for \qty{7320} instances of
    the Gordon--Salmond--Smith system with $\upsilon \sim \mathcal{N}(0, 3^2)$
    and $\nu \sim \mathcal{U}(-0.25, 0.25)$. A point (X, Y) on the graph
    corresponds to Y\% of errors being smaller than X.
    UxHw consistently achieves lower errors than the respective equal-speed
    Monte Carlo. Average \textbf{(B)} RMSE and \textbf{(C)} effective sample
    size for $\upsilon \sim \mathcal{N}(0, 3^2)$ and $\nu \sim
    \mathcal{U}(-0.05, 0.05)$ for different particle counts.}
    \label{FigRmseEssParticleCount}
\end{figure}

UxHw improves accuracy and enables more efficient use of particles compared to
Monte Carlo. Figure~\ref{FigRmseEssParticleCount} shows (\textsc{\large b}) the
average RMSE (lower is better) and (\textsc{\large c}) the average effective
sample size~\cite{kong1994sequential} (higher is better, max one) for 100 filter
trials using the UxHw approach and the Monte Carlo alternative with the same
execution time, for $\upsilon \sim \mathcal{N}(0, 3^2)$ and $\nu \sim
\mathcal{U}(-0.05, 0.05)$. On average, the UxHw approach has 3.3\% lower RMSE
and is better than the Monte Carlo alternative in 11 out of 16 UxHw $\eta$ and
particle count configurations. For 1000 particles, on average the UxHw 8
approach achieves a 9\% lower RMSE compared to the Monte Carlo alternative.
While UxHw trials with bigger $\eta$ remain competitive, they do not offer
consistent advantage at this filter size. For filters with fewer particles, UxHw
16 achieves as much as 5\% lower RMSE (400 particles), while UxHw 8 achieves as
much as 18.9\% lower RMSE (600 particles).

\subsubsection*{Likelihood evaluation speed}

\begin{figure}[t]
    \centering
    \includegraphics[clip,trim={0.1cm 0.2cm 0.1cm 0.9cm},width=0.66\columnwidth,alt={Picture of UxHw-FPGA-17k commercial system-on-module}]{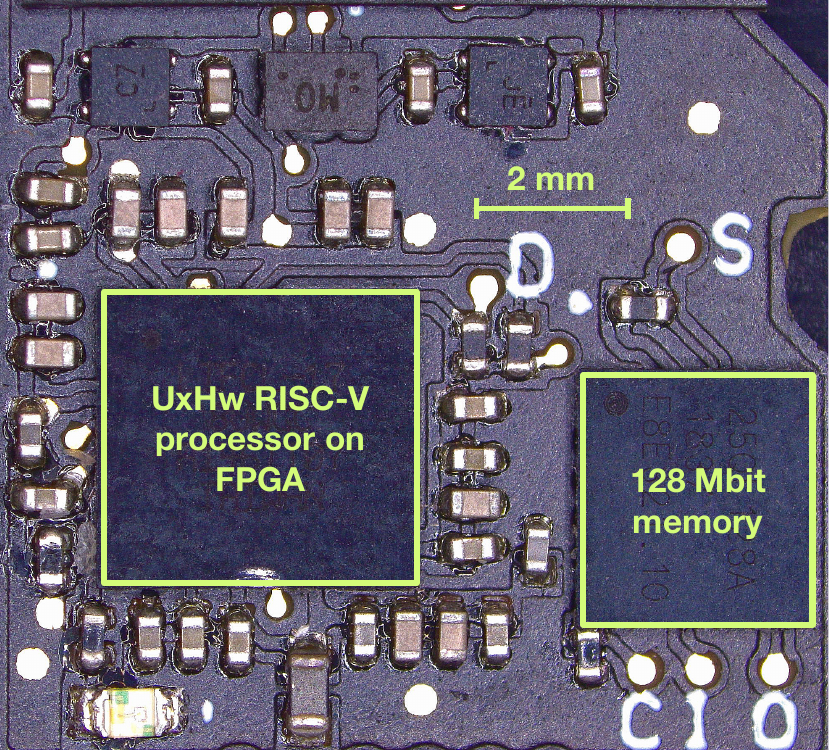}
    \caption{Commercially-available system-on-module UxHw-FPGA-17k for native
    uncertainty tracking with \qty{45}{\mega\hertz} clock speed,
    \qty{320}{\kibi\byte} RAM, \qty{99}{\milli\watt} base power,
    \qty{10}{\milli\meter}~$\times$~\qty{40}{\milli\meter} size, based on RISC-V
    RV32IM.}
    \label{fig:NativeUncertaintyTrackingHardware}
\end{figure}

UxHw achieves faster computation of the predictive-lookahead likelihood compared
to the Monte Carlo approach. Figure~\ref{fig:NativeUncertaintyTrackingHardware}
shows a photo of UxHw-FPGA-17k, the platform for native uncertainty-tracking
where we run the UxHw and Monte Carlo variants of the filters to benchmark their
speed. Table~\ref{TableUxHwEqMCp99Values} shows the latency of computing the
likelihood with the given $\eta$, and the Monte Carlo iteration count necessary
to achieve equal or lower absolute error with 99\% confidence (EqMCp99). UxHw
8 achieves as much as 37.7$\times$ speedup over the respective EqMCp99
(MC 701). As UxHw $\eta$ grows, the UxHw approach shows diminishing
returns in same-accuracy-EqMCp99 speedup: as much as 7.9$\times$ speedup for
UxHw 64 against MC \qty{11046}{}.

\edef\tempa{}%

\begin{table}[t]
\centering
\caption{Column \emph{EqMCp99} shows the minimum amount of Monte Carlo
iterations necessary to yield result better than the corresponding UxHw tuning
at least 99\% of the time. \emph{Latency} columns show speeds on FPGA-17k. UxHw
achieves as much as 37.7$\times$ speedup over the corresponding 99\%-confidence
equal-accuracy Monte Carlo (same hardware). UxHw configurations with $\eta
> 64$ do not run on FPGA-17k due to processor RAM limits.}
\label{TableUxHwEqMCp99Values}
\begin{tabular}{S c|S S||c}
\hline
\multicolumn{2}{c|}{\textbf{UxHw} (FPGA-17k)} & \multicolumn{2}{c||}{\textbf{Monte Carlo} (FPGA-17k)} & \textbf{Speedup}\\
\hline
\textbf{Latency (ms)} & \textbf{\(\eta\)} & \textbf{EqMCp99} & \textbf{Latency (ms)} &  \\ 
\hline
\rowcolor{a}   10       &   8   &   701  & 377    & 37.7$\times$        \\  
\rowcolor{b}   41       &   16  &  1605  & 862    & 21.0$\times$        \\  
\rowcolor{a}  171       &   32  &  3388  & 1820   & 10.6$\times$         \\  
\rowcolor{b}  750       &   64  & 11046  & 5935   & 7.9$\times$         \\  
\if\relax\detokenize\expandafter{\tempa}\relax
\else
\rowcolor{a}   &    &   &    &   \\  
\rowcolor{b}   &    &   &   &   \\   
\fi
\hline
\multicolumn{5}{p{0.9\columnwidth}}{\scriptsize EqMCp99 values are from
empirical datasets each with \qty{6089755} samples with
$\upsilon\sim\mathcal{N}(0, 3^2)$ and $\nu \sim \mathcal{U}(-0.25,0.25)$.
}
\end{tabular}
\end{table}

\subsubsection*{Robustness}

\begin{figure}[tb]
    \centering
    \begin{subfigure}[t]{0.54\columnwidth}
        \centering
        \includegraphics[trim={0.25cm 0.25cm 0.3cm 0.25cm},clip,width=\textwidth]{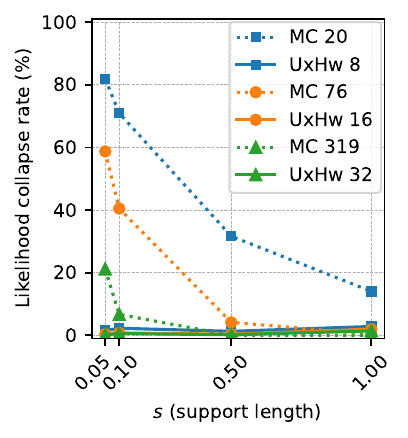}
        \caption{}
        \label{FigFalseZeroes}
    \end{subfigure}
    \hfill
    \begin{subfigure}[t]{0.44\columnwidth}
        \centering
        \includegraphics[trim={0.15cm 0.25cm 0.24cm 0.25cm},clip,width=\textwidth]{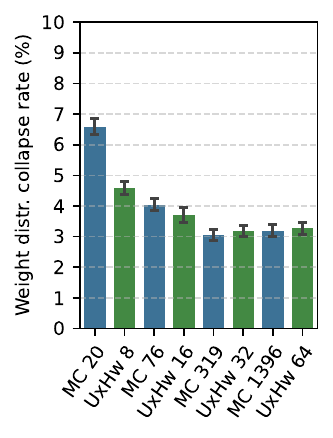}
        \caption{}
        \label{FigWeightCollapseRate}
    \end{subfigure}
    \caption{\textbf{(a)} Likelihood collapse rate for three UxHw and
equal-speed Monte Carlo variants, with $\upsilon \sim \mathcal{N}(0, 3^2)$ and $\nu \sim
\mathcal{U}(-\frac{s}{2},\frac{s}{2})$. UxHw lowers the collapse rate by as much as 80
percentage points ($s=0.05$). \textbf{(b)} Filter weight distribution collapse rate (due to failure to explain the observation) for $s=0.1$. With UxHw the
weight distribution collapses less frequently by as much as 2 percentage points (UxHw~8).}
    \label{FigCombined}
\end{figure}

Because the UxHw approach does not rely on random sampling, it is more
robust to noise outliers compared to Monte Carlo for systems with narrow
observation noise models.
Because Monte Carlo samples the transition model and then evaluates the
observation noise density (Figure~\ref{LykProxyComputationGraph}a), outlier
transitions and observations coinciding with a bounded observation model lead to
significant probability $p$ for the Monte Carlo sampling to miss the
observation support, yield zero, and not inform about the likelihood (other
than scaling). As $p$ grows, there is increasing chance that all iterations of
the same evaluation yield zero whereby the likelihood collapses to zero. This phenomenon is more frequent in
low-iteration Monte Carlo evaluation schemes which are also the fastest, and
thus most desirable for the particle filter~\cite{blanco2010optimal}.
Figure~\ref{FigFalseZeroes} shows the likelihood collapse rate for UxHw and respective equal-speed
Monte Carlo variants. For uniform observation noise with scale $0.05$, the
filter falsely evaluates the likelihood to zero $81.89\%$ of the time (MC 20),
wasting computing resources, whereas UxHw~8 has $1.52\%$ false-zero rate.
Supplementary Figure~\suppref{FigFalseZeroesDetail}{7} shows a detail version of
Figure~\ref{FigFalseZeroes} focusing on small collapse rates.
Figure~\ref{FigWeightCollapseRate} shows the rate of weight distribution
collapses in the filter (necessitating filter reset). With UxHw 8 the weight
distribution collapses less frequently by 2 percentage points.
Supplementary Tables~\suppref{TableLikelihoodCollapseRates}{II}
and~\suppref{TableFilterCollapsePercent}{III} list the data of
Figure~\ref{FigCombined}.

%% file: text/conclusion.tex
\section{Conclusions}

The UxHw approach shows better performance than Monte Carlo at computing the
likelihood in systems with broad transition noise and narrow bounded observation
noise. In such scenarios, relative to an equal-accuracy Monte Carlo baseline,
UxHw 8 achieves as much as 37.7$\times$ speedup when observation noise is
narrow and uniform, while also, relative to an equal-speed Monte Carlo
baseline UxHw, improving RMSE by 3\% on average (and by as much as 18.9\%
on UxHw 8 with 600 particles). As UxHw $\eta$ grows, the UxHw
approach shows diminishing returns in same-accuracy speedup. As particle count
grows, the UxHw-based filters make better use of each particle, increasing
effective sample size. The method improves the speed--fidelity tradeoff and
lowers the likelihood collapse rate by as much
as 80 percentage points on UxHw 8. This increased likelihood
robustness translates to fewer filter weight distribution collapses by 2
percentage points for UxHw 8 and 0.30 percentage points for UxHw 16. UxHw
employs deterministic computation and thus avoids the high variance common in
Monte Carlo methods.

%% file: text/supplementary.tex
\section{Supplementary Material}

\balance

\subsection{Particle filtering recap}
\label{SuppSectionParticleFilteringRecap}

The particle filter algorithm, also known as Sequential Importance Sampling with
Resampling (SIR), estimates the state of nonlinear, non-Gaussian dynamic systems
by maintaining a set of weighted samples (particles).

Let $\mathbf{x}_k$ denote the state at time $k$, $\mathbf{u}_k$ the control
input, and $\mathbf{z}_k$ the observation.
Equations~\ref{eq:ParticleFilterProcessModel}
and~\ref{eq:ParticleFilterObservationModel} represent the state-space model of
a Markov system.
\begin{align}
    \mathbf{x}_k &\sim p(\mathbf{x}_k \mid \mathbf{x}_{k-1}, \mathbf{u}_k), \label{eq:ParticleFilterProcessModel} \\
    \mathbf{z}_k &\sim p(\mathbf{z}_k \mid \mathbf{x}_k). \label{eq:ParticleFilterObservationModel}
\end{align}
The particle filter approximates the posterior distribution $p(\mathbf{x}_k \mid
\mathbf{x}_{1:k-1}, \mathbf{z}_{1:k}, \mathbf{u}_{1:k})$ using a set of $N$ particles
$\{\mathbf{x}_k^{(i)}, w_k^{(i)}\}_{i=1}^N$, where each particle
$\mathbf{x}_k^{(i)}$ has an associated weight $w_k^{(i)}$.
The particle filter algorithm is:
\begin{enumerate}
    \item \textbf{Initialization:} Draw $N$ particles from the initial
    distribution $p(\mathbf{x}_0)$ and set uniform weights $w_0^{(i)} = 1/N$.
    \item \textbf{For each time step $k$:}
    \begin{enumerate}
        \item \textbf{Prediction:} For each particle, sample a new state:
        \begin{equation}
            \mathbf{x}_k^{(i)} \sim p(\mathbf{x}_k \mid \mathbf{x}_{k-1}^{(i)}, \mathbf{u}_k).
        \end{equation}
        \item \textbf{Update:} Compute the importance weight for each particle
        based on the likelihood of the new observation and the proposal
        distribution. In general, the importance weight is given by
        \begin{equation}
            w_k^{(i)} \propto w_{k-1}^{(i)} \cdot \frac{p(\mathbf{z}_k \mid \mathbf{x}_k^{(i)}) \, p(\mathbf{x}_k^{(i)} \mid \mathbf{x}_{k-1}^{(i)}, \mathbf{u}_k)}{q(\mathbf{x}_k^{(i)} \mid \mathbf{x}_{k-1}^{(i)}, \mathbf{z}_k, \mathbf{u}_k)}, \label{eq:ParticleFilterUpdateGeneric}
        \end{equation}
        where $q(\cdot)$ is the proposal distribution density. In the standard
        \emph{bootstrap} particle filter, the proposal is chosen as the
        transition model density, i.e., $q(\mathbf{x}_k \mid \mathbf{x}_{k-1},
        \mathbf{z}_k, \mathbf{u}_k) = p(\mathbf{x}_k \mid \mathbf{x}_{k-1},
        \mathbf{u}_k)$, which simplifies the weight update to
        \begin{equation}
            w_k^{(i)} \propto w_{k-1}^{(i)} \cdot p(\mathbf{z}_k \mid \mathbf{x}_k^{(i)}).   \label{EquationParticleFilterResamplingGeneric}
        \end{equation}
        Then, normalize the weights so that $\sum_{i=1}^N w_k^{(i)} = 1$.
        \item \textbf{Resampling:} To avoid degeneracy (where all but one
        particle have negligible weight), resample the particles according to
        their weights to obtain a new set of equally weighted particles.
    \end{enumerate}
\end{enumerate}
To optimize for speed, different particle filter variants run the resampling
step on specific conditions. Without loss of generality and for clarity of
exposition, we always perform the resampling step.

This letter presents a method to achieve the update step
(Equation~\ref{eq:ParticleFilterUpdateGeneric}) and avoiding the degeneracy that
the resampling step (Equation~\ref{EquationParticleFilterResamplingGeneric})
attempts to fix. Our method uses recently-developed techniques for efficient
representation, arithmetic, integration, and sampling of probability
distribution, in computation.

\begin{figure*}[htb]
\centering
\includegraphics[trim={0.4cm 0cm 10.5cm 0cm},clip,width=0.99\textwidth]{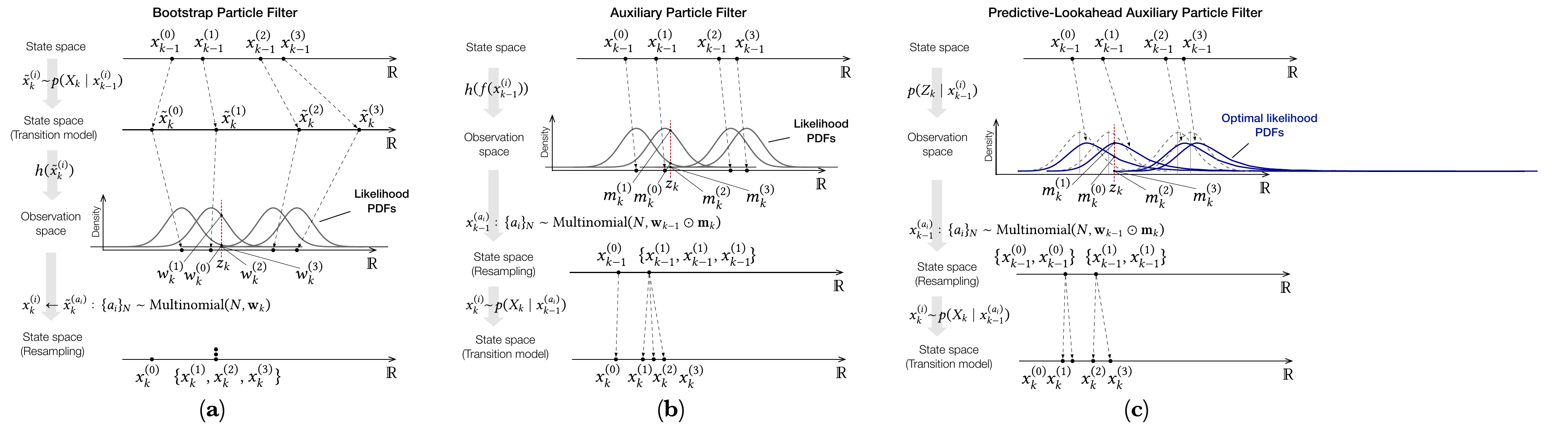}
\caption{The standard bootstrap particle filter, the auxiliary particle filter,
and the predictive-lookahead auxiliary particle filter. \textbf{(a)} The
standard bootstrap particle filter evaluates the likelihood density at the
measurement to compute the particle weights which control the probability of
choosing the given particle at the resampling step. \textbf{(b)} To reduce
wasted work of propagating ultimately unlikely particles, the auxiliary particle
filter computes the proxy likelihoods $m_k^{(i)}$ before sampling the noisy
transition model. \textbf{(c)} The predictive-lookahead variant computes
$m_k^{(i)}$ as the composite uncertainty of both the transition and observation
models, rather than just the observation model, making the filter more robust to
measurement outliers and more compatible with low-variance observation models.}
\label{FigParticleFilterVariantsAuxiliary}
\end{figure*}

Figure~\ref{FigParticleFilterVariantsAuxiliary} shows the differences between
the bootstrap particle filter, the auxiliary particle filter, and the
predictive-lookahead auxiliary particle filter. Both auxiliary variants use the
optimal proposal density for the importance sampling step.

\subsection{Likelihood evaluation and benchmarking for the Gordon--Salmond--Smith system}
\label{SecLikelihoodBenchmarkingDetails}

Our hypothesis targets systems where the transition noise scale $s_\upsilon$ is
larger than the observation noise scale $s_\nu$. We benchmark with three
parametric distribution families for the additive transition noise distribution
$\mu_\upsilon(0, s_\upsilon)$ and the additive observation noise distribution
$\mu_\nu(0, s_\nu)$; the first parameter is location and the second is scale.
We benchmark for the product combinations of the following values:
\begin{align}
    \mu_\upsilon    &   \in[\mathrm{Gaussian}, \mathrm{Laplacian}, \mathrm{Uniform}],\\
    \mu_\nu         &   \in[\mathrm{Gaussian}, \mathrm{Laplacian}, \mathrm{Uniform}],\\
    s_\upsilon      &   \in[3, 1, 0.5],\\
    s_\nu           &   \in[1, 0.5, 0.1, 0.05].
\end{align}
This tallies to 108 system parameters configurations.\footnote{The improvement
hypothesis for the UxHw approach assumes combinations where $s_\upsilon>s_\nu$.
We use the results from other combinations to provide further context of the
UxHw method for systems where the assumption does not hold.} The scale parameter
corresponds to the standard deviation for the Gaussian, the scale for the
Laplacian, and the distribution support length for the Uniform.

For benchmarking the computation of the likelihood itself we pick 61
linearly-spaced locations in the range $[-30, 30]$ where the
Gordon--Salmond--Smith process takes most state values.

For each combinatorial choice of $s_\upsilon$ and $s_\nu$, and for each location
$x$ as above, we produce simulated measurements with predetermined error
$\epsilon$, with
\begin{equation}
    \epsilon\in[\pm1, \pm0.5, \pm0.1, \pm0.05],
\end{equation}
to benchmark the accuracy of the approach for different probable measurements
off the true state, using the formula $h(f(x)) - \epsilon$.

The Monte Carlo estimator for the proxy likelihood $\tilde{m}_{k,M}^{(i)}$ is:
\begin{equation}
    \tilde{m}_{k,M}^{(i)} = \tilde{p}_M({z}_k \mid {x}_{k-1}^{(i)}) = \frac{1}{M} \sum_{j=1}^M p({z}_k \mid x_j),\hspace{1ex} x_j \stackrel{\text{iid}}{\sim} p(X_k \mid {x}_{k-1}^{(i)}).
\end{equation}
To benchmark the speedup of UxHw with $\eta\in[8, 16, 32, 64]$, against the
corresponding p99-equal-accuracy Monte Carlo alternatives, we compute with:
\begin{equation}
    M\in[701, 1605, 3388, 11046].
\end{equation}
To benchmark the accuracy improvement of UxHw against the corresponding
equal-speed Monte Carlo alternatives, we compute with:
\begin{equation}
    M\in[20, 76, 319, 1396].
\end{equation}

 For each $M$, we repeat 1000 times to capture the variance of the
approach. The Monte Carlo approach converges to the true density distribution:
for each system configuration and for each $\mathrm{X}$ and $\mathrm{SimErr}$ we
compute the ground truth value of the likelihood using Monte Carlo evaluation
with $M_{\mathrm{gt}}=10^7$ iterations.

The uncertainty representation underpinning UxHw needs $\eta$ to be a power of
two~\cite{bilgin2025quantization}. 
On the native uncertainty-tracking hardware FPGA-17k we compute for the
configuration subset with:
\begin{equation}
    \eta\in[8, 16, 32, 64].
\end{equation}
While other UxHw variants support configuration with larger
$\eta$, FPGA-17k supports up to 64 because of the limited \qty{320}{\kibi\byte}
RAM. We compute average timing by repeating 100 or 50 evaluations in a loop and
averaging the time elapsed, measured via GPIO pin state transitions. (For
constant size, each method runs in constant time. For Monte Carlo, each run
leads to a different result.)

Listing~\ref{ListingComplementaryCode} contains example function
definitions complementary to the source code of
Figure~\ref{FigCodeComparePointwiseMonteCarloUxHw}
Listings~\ref{ListingProxyLyksPointwiseApproach}--\ref{ListingProxyLyksUxHwMarginalApproach}
for the Gordon--Salmond--Smith system and for the configuration with $\upsilon
\sim \mathcal{N}(0, 3^2)$ and $\nu \sim \mathcal{U}(-0.25,0.25)$. When compiling
for UxHw, functions \codelight{UxHwDoubleGaussDist} and
\codelight{UxHwDoubleUniformDist} are not C implementations but rather
thinly-wrapped UxHw-microarchitecture instructions for assigning parametric
distribution to variables~\cite{tsoutsouras2022laplace, petangoda2025monte}.
When compiling for non-UxHw they are samplers from the GNU Scientific Library.

\begin{figure}[htb]
\centering
\begin{lstlisting}[language=C,style=c,caption={Example C function definitions for a configuration of
the Gordon--Salmond--Smith system and with $\upsilon_k \sim \mathcal{N}(0, 3^2)$
and $\nu_k \sim \mathcal{U}(-0.25,0.25)$.},label=ListingComplementaryCode]
double transition_model_ideal(double x, double t) {
	return 0.5 * x + 25 * x / (1 + pow(x, 2)) + 8 * cos(1.2 * t);
}

double observation_model_ideal(double x) {
	return pow(x, 2) / 20;
}

double transition_noise_model() {
	const double	u_scale = 3;	// (*@System with $\upsilon_k \sim \mathcal{N}(0, 3^2)$@*)
	return UxHwDoubleGaussDist(0, u_scale);
}

double observation_noise_model() {
	const double	nu_scale = 0.5;	// (*@System with $\nu_k \sim \mathcal{U}(-0.25,0.25)$@*)
	return UxHwDoubleUniformDist(-nu_scale/2, nu_scale/2);
}
\end{lstlisting}
\label{FigComplementaryCodeListings}
\end{figure}

\begin{figure}[htb]
    \centering
    \includegraphics[width=0.9\columnwidth]{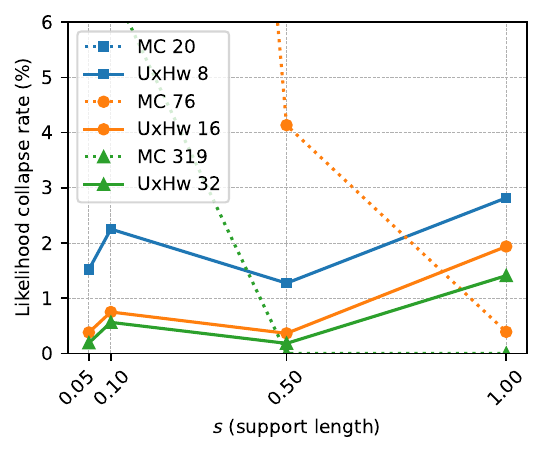}
    \caption{Detail of likelihood collapse rate
(Figure~\ref{FigFalseZeroes}) for rates 0\% to 6\% for three UxHw and
equal-speed Monte Carlo variants, with $\upsilon \sim \mathcal{N}(0, 3^2)$ and
$\nu \sim \mathcal{U}(-\frac{s}{2},\frac{s}{2})$.}
    \label{FigFalseZeroesDetail}
\end{figure}

Figure~\ref{FigFalseZeroesDetail} shows a zoomed-in detail version of
Figure~\ref{FigFalseZeroes} focusing on likelihood collapse rates 0\%--6\%, for
the system with $\upsilon \sim \mathcal{N}(0, 3^2)$.
Table~\ref{TableLikelihoodCollapseRates} lists the data of
Figure~\ref{FigFalseZeroes}. Because the UxHw approach does not rely on random
sampling, it achieves higher robustness relative to the equal-speed Monte Carlo
approach for small observation noise scales $s_\nu$. Also, because the
computation latency on the UxHw-FPGA-17k follows a power-law relationship with
$\eta$, bigger UxHw $\eta$ configurations present diminishing returns over the
corresponding equal-speed Monte Carlo.

Table~\ref{TableFilterCollapsePercent} lists the data of the weight
distribution collapse rate for Figure~\ref{FigWeightCollapseRate}. In the UxHw
approach, improvements to the likelihood collapse rate translate to moderate
improvements in the weight distribution and thus to the robustness of the
filter.

\subsubsection*{Variance-based alternative (VarSum)}

A simple and performant alternative for approximating the lookahead likelihood
$p(z_t \mid x_{t-1}^{(i)})$ is an EKF-style
linearization~\cite{johansen2009tutorial} using the Jacobian matrix of the
observation model. Equation~\ref{eq:PredictiveLookaheadLikelihoodLinearization}
shows the Gaussian approximation of the likelihood density using this approach.
\begin{equation}
    \hat{p}(z_t \mid x_{t-1}^{(i)}) \approx \mathcal{N}\!\big( z_t;\,h(f(x_{t-1}^{(i)})),\; (h'(f(x_{t-1}^{(i)})))^2s_\upsilon^2 + s_\nu^2 \big). \label{eq:PredictiveLookaheadLikelihoodLinearization}
\end{equation}
In the benchmarks of Section~\ref{Section_Results}, VarSum implements an
adaptive approach of evaluating a Gaussian approximation or Laplacian
approximation density.

\subsubsection*{Root-mean-square error (RMSE)}

The root-mean-square error is a standard measure of estimation accuracy. It
quantifies the average magnitude of the estimation error between the true latent
state and the filter estimate across all time steps, penalizing larger errors
more heavily due to the squaring operation. Let $x_k$ denote the true state and
$\hat{x}_k$ the estimated state at time step $k$. Equation~\ref{EqRMSE} shows
the formula for RMSE computed over $T$ time steps:
\begin{equation}
\mathrm{RMSE} = \sqrt{\frac{1}{T}\sum_{t=1}^{T}(x_k - \hat{x}_k)^2}. \label{EqRMSE}
\end{equation}
Lower RMSE indicates closer agreement between the estimated and true state
trajectories, while a higher RMSE suggests poor tracking performance or filter
divergence.

\subsubsection*{Effective sample size (ESS)}

The effective sample size quantifies the number of particles that contribute
meaningfully to the posterior estimate after normalization of the particle
weights~\cite{kong1994sequential}. Small ESS indicates particle degeneracy,
where few particles dominate the approximation, while a large ESS implies a
diverse and well-represented particle set. For normalized weights ${w_i}$,
Equation~\ref{EqESS} shows the formula for ESS:
\begin{equation}
\mathrm{ESS} = \frac{1}{\sum_{1}^{N} w_i^{2}}. \label{EqESS}
\end{equation}
Since we compute ESS inside the benchmarks at every particle filter time step
(Section~\ref{Section_Results}), for the C implementation we prefer a
numerically stable normalized variant which behaves better in the presence of
small minor normalization errors due to floating-point rounding: $\mathrm{ESS} =
(\sum_i w_i)^2 / \sum_i w_i^2$,

\begin{table}[htb]
\centering
\caption{Detailed results dataset for likelihood collapse rate (\%) of
Figure~\ref{FigFalseZeroes}. Column \emph{Equal-Speed MC} reports the collapse
rate for the smallest Monte Carlo simulation size whose runtime no faster than
the corresponding UxHw tuning.}
\label{TableLikelihoodCollapseRates}
\begin{tabular}{r S|S S|S}
\hline
\multicolumn{2}{c|}{} & \multicolumn{2}{c|}{\textbf{Likelihood collapse rate (\%)}}\\
\cline{3-4}
\textbf{\(s_\nu\)} & \textbf{UxHw \(\eta\)} & \textbf{UxHw} & \textbf{Equal-Speed MC} \\
\hline
\rowcolor{a} 0.05 & 8  & 1.5209 & 81.8912 \\
\rowcolor{b} 0.05 & 16 & 0.3802 & 58.7068 \\
\rowcolor{a} 0.05 & 32 & 0.1901 & 21.1480 \\
\rowcolor{b} 0.05 & 64 & 0.0000 &  0.5511 \\
\hline
\rowcolor{a} 0.10 & 8  & 2.2514 & 70.9223 \\
\rowcolor{b} 0.10 & 16 & 0.7504 & 40.4857 \\
\rowcolor{a} 0.10 & 32 & 0.5628 &  6.6705 \\
\rowcolor{b} 0.10 & 64 & 0.5628 &  0.0093 \\
\hline
\rowcolor{a} 0.50 & 8  & 1.2727 & 31.6281 \\
\rowcolor{b} 0.50 & 16 & 0.3636 &  4.1343 \\
\rowcolor{a} 0.50 & 32 & 0.1818 &  0.0016 \\
\rowcolor{b} 0.50 & 64 & 0.1818 &  0.0000 \\
\hline
\rowcolor{a} 1.00 & 8  & 2.8169 & 14.0448 \\
\rowcolor{b} 1.00 & 16 & 1.9366 &  0.3906 \\
\rowcolor{a} 1.00 & 32 & 1.4084 &  0.0003 \\
\rowcolor{b} 1.00 & 64 & 0.8802 &  0.0000 \\
\hline
\end{tabular}
\end{table}

\begin{table}[htb]
\centering
\caption{Detailed results dataset for weight distribution collapse rate (\%) of
Figure~\ref{FigWeightCollapseRate}. Data for configuration with $s_\nu=0.1$.}
\label{TableFilterCollapsePercent}
\begin{tabular}{S|S S}
\hline
& \multicolumn{2}{c}{\textbf{Weight distrbution collapse rate (\%)}} \\
\hline
\textbf{UxHw \(\eta\)} & \textbf{UxHw} & \textbf{Equal-Speed MC} \\
\hline
\rowcolor{a} 8  & 4.59 & 6.59 \\
\rowcolor{b} 16 & 3.70 & 4.04 \\
\rowcolor{a} 32 & 3.18 & 3.05 \\
\rowcolor{b} 64 & 3.27 & 3.20 \\
\hline
\end{tabular}
\end{table}